\documentclass{pnastwo_preprint}

\pdfoutput=1

\usepackage{amsmath, qtree, graphicx}

\newcommand{\psff}[1]{\includegraphics{#1.pdf}}

\usepackage{nicefrac}
\usepackage{tikz}
\newsavebox{\agent}
\savebox{\agent}{
  \begin{tikzpicture}[
      scale=0.07,
      every node/.style={coordinate, draw}
    ]

    \node[circle, inner sep = 0pt, minimum size = 6pt] (head) at (0, 0) {};
    \node (middle) at (0, -3.5) {};
    \node (bottom) at (0, -5) {};
    \node (left hand) at (-3, -2.5) {};
    \node (right hand) at (3, -2.5) {};
    \node (left foot) at (-2.5, -8) {};
    \node (right foot) at (2.5, -8) {};

    \path (head) edge (middle) edge (bottom);
    \path (middle) edge (left hand) edge (right hand);
    \path (bottom) edge (left foot) edge (right foot);
  \end{tikzpicture}
}

\begin{document}

\title{The entropic basis of collective behaviour}

\author{Richard P Mann\affil{1}{Swiss Federal Institute of Technology, Z\"{u}rich,
Switzerland}\affil{2}{Department of Mathematics, Uppsala University, Uppsala, Sweden}
\and
Roman Garnett\affil{3}{Department of Computer Science and Engineering, Washington University in St.\ Louis, St.\ Louis, MO, United States of America}}



\maketitle

\begin{article}
\begin{abstract}
{In this paper, we identify a radically new viewpoint on the collective behaviour of groups of intelligent agents. We first develop a highly general abstract model for the possible future lives that these agents may encounter as a result of their decisions. In the context of these possible futures, we show that the \emph{causal entropic principle,} whereby agents follow behavioural rules that maximise their entropy over all paths through the future, predicts many of the observed features of social interactions between individuals in both human and animal groups. Our results indicate that agents are often able to maximise their future path entropy by remaining cohesive as a group, and that this cohesion leads to collectively intelligent outcomes that depend strongly on the distribution of the number of future paths that are possible. We derive social interaction rules that are consistent with maximum-entropy group behaviour for both discrete and continuous decision spaces. Our analysis further predicts that social interactions are likely to be fundamentally based on Weber's law of response to proportional stimuli, supporting many studies that find a neurological basis for this stimulus-response mechanism, and providing a novel basis for the common assumption of linearly additive `social forces' in simulation studies of collective behaviour. }
\end{abstract}



\dropcap{C}ollective decision making and the emergence of collective intelligence are key areas of study in the fields of animal behaviour and social science. Since Francis Galton observed the power of the central limit theorem to provide an accurate estimate for the weight of a bull by averaging individual opinions (as told by James Surowiecki \cite{surowiecki2005wisdom}), the ability of groups to make decisions that improve on the accuracy of the individuals comprising them has continued to surprise researchers. Human \cite{Woolley2010efa}, animal \cite{sumpter2010cab}, and even algorithmic \cite{rokach2010ebc} groups have been shown to improve on individual performance in estimation problems (Galton's bull example), navigation \cite{simons2004mwt}, identifying superior options \cite{seeley2010hd}, and prediction tasks \cite{wolfers2006pmi}. In an age of unprecedented global connectivity of individuals through web and mobile internet technologies, the opportunity to understand the origins of social behaviour and leverage collective intelligence is greater than ever before.

Much is already known about how the transfer of information by individuals can lead to intelligent outcomes on the level of the group. Models of social contagion \cite{Salganik:2006mi,Hedstrom:2008qa, mann2013tdo}, quorum decision making \cite{sumpter2008cdm, Ward2008ly, ward2012quo}, Bayesian social decision rules \cite{perez2011cab, arganda2013acr}, and information cascades \cite{easley2010nca, mann2013humbugs} all provide a detailed theory for how each agent in a group can acquire and use information from other individuals' actions, and under what conditions this leads to improved or disrupted decision making by those individuals.

However, when we face the challenge of understanding the collective behaviour of the millions of connected individuals now on our planet, the prospect of beginning that process at the level of a single individual decision maker is daunting. Statistical mechanics, and particularly the principle of maximum entropy, provide an expedient methodology for studying the behaviour of large systems with many interacting elements. More recently, the principle of maximum entropy production (reviewed by \cite{niven2009sso}) has enabled these methods to be applied to more-general flow systems outside of the classical notion of equilibrium, and causal entropy \cite{wissner2013cef} has been proposed to extend this to the case where the individual elements of the system exhibit intelligence.

Statistical mechanics has already provided a fruitful route to understanding collective behaviour, in particular collective motion, via the abstraction of \emph{social forces}: pseudo-forces that can modify an agent's energy depending on its alignment with or proximity to other individuals \cite{vicsek1995nto}, or explicitly provide a physical force to alter the agents' motion \cite{couzin2002cma, helbing1995sfm}. Such approaches have been able to demonstrate why human and animal groups undergo phase transitions between different quasi-equilibria in analogy with the phase transitions seen in statistical-mechanical systems, and have been developed to a particularly high degree of sophistication in the study of human crowds \cite{helbing2009pca}, where they are used to understand disasters such as at Hillborough (1989) and the Love Parade (2010) \cite{helbing2012dca}. However, social forces are a convenient abstraction of psychological choices, and therefore are typically adjusted to fit observations, rather than being based on the fundamental logic of interactions.

In this paper we demonstrate a new way to understand collective behaviour, from a purely entropic viewpoint, without any specification of social information transfer, social forces, or individual interaction rules. We do this by building on the causal entropic framework of \cite{wissner2013cef}. By specifying our uncertainty about the long-term futures of a group of agents, we will show that the decisions this group make \emph{now} can be predicted. We will show that the social rules of interaction, and the social forces assumed in many studies of collective behaviour emerge not from any consideration of the adaptiveness of the agents' choices, nor from any consideration of their immediate needs or desires, but simply from a tacit assumption that their long-term actions are maximally uncertain.

\section{The causal entropic principle}
The causal entropic principle is an assertion about our knowledge of a system's future path through state space. This is fundamentally an argument from a principle of maximum \emph{ignorance} -- we deem ourselves to be as uncertain as possible about the path an agent will take through all the future options available. As we shall show, this counter-intuitively provides us with information about which choices the agent is likely to make now. In previous work, Wissner-Gross \& Freer \cite{wissner2013cef} derived a `causal entropic force' that drives systems towards locally available new microstates that permit a greater number of available paths through future state space. In the cases presented by \cite{wissner2013cef}, this force acts upon particles moving in a continuous, bounded Euclidean space. As ergodic principle for equilibria states that any microstate of the particles in the gas is equally probable, so in a causal-entropic system all available future paths are assigned an equal probability. Therefore the probability of any new reachable microstate being selected is proportional to the number of future state-space paths that it makes available. This causal entropic force was shown to cause a diverse range of systems to behave in apparently intelligent ways, mimicking for example animal use of tools, or complex co-operation. Inspired by these examples, we consider whether the same principle can predict the interactions between individuals in groups that are the foundation of collective cognition and intelligence.

\subsection{Application to collective decisions: A toy example}
Consider a group of agents who must decide between two options, A and B. In this example, the information about the future is the following: behind one door lies four more options; behind the other, there is only one. This is illustrated in Figure 1. Assuming that the door with four options is equally likely to be either A or B, what distribution of the agents between the two options will maximise their expected entropy, over the possible future paths to the final level of the branching tree?

\tikzset{
  empty/.style  = {draw, font={\footnotesize\sffamily}},
  filled/.style = {fill = red!60!white}
}

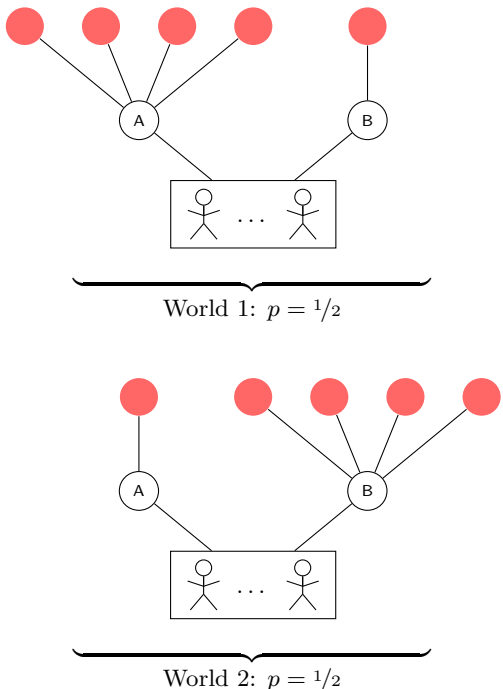
\begin{figure}
  \hspace*{0.5cm}
  \begin{tikzpicture}[
      every node/.style = {circle, minimum size = 5mm},
      grow = up,
      level 1/.style = {sibling distance = 3cm},
      level 2/.style = {sibling distance = 1cm},
      level distance = 1.25cm
    ]
    \node [draw, rectangle] {\usebox{\agent} \raisebox{1.75ex}{\dots} \usebox{\agent}}
    child {
      node [empty] {B}
      child { node [filled] {} }
    }
    child {
      node [empty] {A}
      child { node [filled] {} }
      child { node [filled] {} }
      child { node [filled] {} }
      child { node [filled] {} }
    };
  \end{tikzpicture}
  \\
  \hspace*{1.7cm}
  $\underbrace{\mspace{260mu}}_{\displaystyle \text{World 1: }p = \nicefrac{1}{2}}$
  \vspace{5ex}
  \\
  \hspace*{2.3cm}
  \begin{tikzpicture}[
      every node/.style = {circle, minimum size = 5mm},
      grow = up,
      level 1/.style = {sibling distance = 3cm},
      level 2/.style = {sibling distance = 1cm},
      level distance = 1.25cm
    ]
    \node [draw, rectangle] {\usebox{\agent} \raisebox{1.75ex}{\dots} \usebox{\agent}}
    child {
      node [empty] {B}
      child { node [filled] {} }
      child { node [filled] {} }
      child { node [filled] {} }
      child { node [filled] {} }
    }
    child {
      node [empty] {A}
      child { node [filled] {} }
    };
  \end{tikzpicture}
  \\
  \hspace*{1.7cm}
  $\underbrace{\mspace{260mu}}_{\displaystyle \text{World 2: }p = \nicefrac{1}{2}}$
\caption{Schematic of a toy example illustrating the causal entropic collective model. A group of agents at the root of the tree must choose between two options: `A' and `B'. Two possible worlds exist: one where option A leads to four more choices and B to one, or one where A leads to one more choice and B to four. The decision rule for the group that maximises their future path entropy averaged over the two possible worlds is a mixture of two binomial distributions, shown in Figure 2.}
\label{fig:schematic1}
\end{figure}

For any given branching tree, entropy is maximised by making any assignment of the agents to each future path that reaches the final level equally probable. Because the graph of choices is a tree, each final option is associated with a single unique path through the future space; therefore, it is equivalent to assign agents randomly to the final nodes on the tree. We aim to find a consistent distribution of agents that maximises the path-entropy over all possible worlds -- a general way for the agents to organise themselves such that their entropy will be as high as possible, on average, in all the worlds they might encounter. Therefore we take each possible tree, weighted by its probability of existing, and assign a uniform multinomial distribution of the agents to its final nodes. We then feed this distribution back to the first choices (in this case, A and B) that the agents must make. Denoting by $N$ the total number of agents, and by $A$ and $B$ the number choosing each door, this model implies that the probability distribution for the number choosing door A is a weighted sum of two binomial distributions, one with $p=\nicefrac{4}{5}$, the other with $p=\nicefrac{1}{5}$. Each has a weight of \nicefrac{1}{2}, since each has a 50\% chance of existing:
\begin{align}
P(A) &= \frac{1}{2} \binom{N}{A} \frac{4}{5}^A\frac{1}{5}^{N-A}  + \frac{1}{2} \binom{N}{A} \frac{1}{5}^A \frac{4}{5}^{N-A} \nonumber \\
&= \frac{1}{2} \binom{N}{A} \frac{4^A + 4^{N-A}}{5^N}.
\label{eqn:toy}
\end{align}
For the case of eight agents picking between these two options, the expected distribution is shown in Figure 2, alongside the distribution we would expect if each agent chose a door independently at random. The exact form of the distribution varies with the total number of agents, as well as with the number of future options. The figure clearly shows that the causal entropic principle, picking randomly from future options rather than the immediately available ones, induces a greater degree of cohesion on the agents -- they are much more likely to choose the same option. This cohesive `force' increases as the difference between the number of options behind each door increases, despite the agents having no information about which door actually contains the greater number of options.
\begin{figure}
\psff{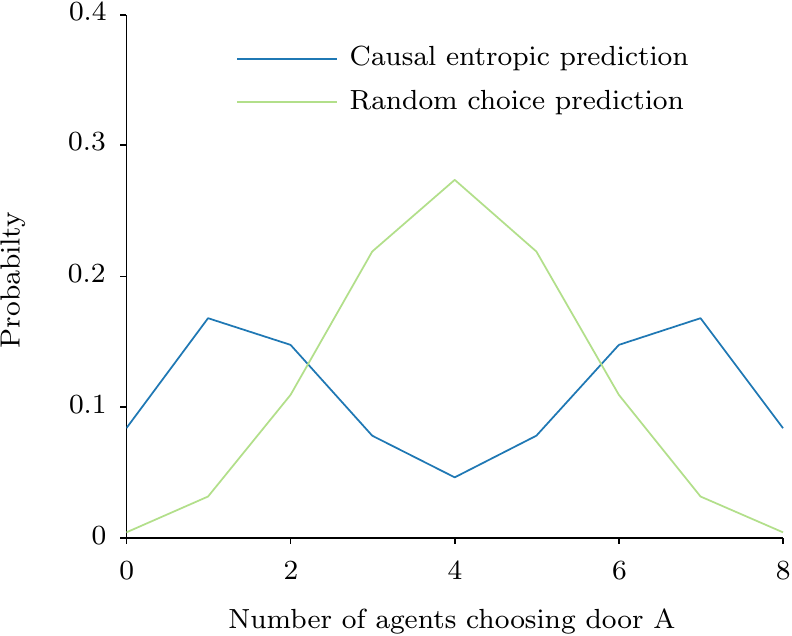}
\caption{An example of predicted decisions by a group of eight agents in a `toy' world: choosing one (unknown) option leads to four possible future paths, and the other to one.  A future path is assigned at random to each agent, averaging over possible configurations of the future world, where the four options may be behind choice A or B. The predicted distribution of agent choices is a weighted sum of binomial distributions, with far greater cohesion than expected if each agent would independently choose a door at random.}
\label{fig:toyexample}
\end{figure}

\section{Collective causal entropic model}
We now expand the toy example above to consider more general collective decisions, where the information about the number of future options is less precise. Letting $P(n_A)$ and $P(n_B)$ describe the probability of finding $n_A$ and $n_B$ future paths behind doors A and B, respectively (assuming for now that these are independent), Equation \ref{eqn:toy} generalises to an infinite sum of probability-weighted binomial distributions.
\begin{align}
P(A \mid N) &= \binom{N}{A} \sum_{n_A=0}^{\infty} \sum_{n_B=0}^{\infty} P(n_A)P(n_B) \frac{n_A^A n_B^{(N-A)}}{(n_A+n_B)^N} \nonumber \\
&= \binom{N}{A} \int_{{R}=0}^1 P(R) R^A (1-R)^{N-A} \, \mathrm{d}R, \label{eqn:general}
\end{align}
where $R =  \frac{n_A}{(n_A+n_B)}$. Clearly the key factor in Equation \ref{eqn:general} that controls the number of agents $A$ choosing door A is the ratio $R$, the proportion of future options that lie behind door A. The problem of estimating the behaviour of the agents is thus largely a problem of estimating $P(R)$, the probability of this ratio.

\subsection{A probability distribution for the number of possible futures}
In general the number of future paths that either A or B may lead to may take any distribution. However, for the purposes of deriving the consequences of a model of collective decision making, we must determine a specific form for $P(n_A)$ and $P(n_B)$, and most importantly for $P(R)$. We propose the following method: confronted with a decision, the agents can represent their belief about the futures available behind each option as a continuing branching tree, in which each branch leads to an unknown number of future choices (illustrated in Figure 3). The number of new choices generated on each branch is determined by some fixed distribution. This is a Galton--Watson (GW) process \cite{watson1875otp}. 


\tikzset{
  start/.style  = {circle, draw, font = {\footnotesize\sffamily}},
  middle/.style = {coordinate},
  filled/.style = {circle, fill = red!60!white,   inner sep = 1.2mm},
  dead/.style   = {circle, fill = black!60!white, inner sep = 1mm}
}

\begin{figure}
  \centering
\begin{tikzpicture}[
    level 1/.style = {sibling distance = 3cm},
    level 2/.style = {sibling distance = 1.5cm},
    level 3/.style = {sibling distance = 1.5cm},
    level 4/.style = {sibling distance = 1.5cm},
    level 5/.style = {sibling distance = 4mm},
    level distance = 1cm,
    grow = up
  ]
  \node [draw, rectangle] {\usebox{\agent} \raisebox{1.75ex}{\dots} \usebox{\agent}}
  child {
    node [start] {B}
    child {
      node [middle] {}
      child {
        node [middle] {}
        child [sibling distance = 1.5cm] {
          node [middle] {}
          child { node [filled] {} }
        }
        child {
          node [middle] {}
          child { node [filled] {} }
          child { node [filled] {} }
          child { node [filled] {} }
          child { node [filled] {} }
        }
      }
      child [sibling distance = 2.5cm] {
        node [middle] {}
        child {
          node [middle] {}
          child { node [filled] {} }
        }
      }
    }
  }
  child {
    node [start] {A}
    child [sibling distance = 1.25cm] {
      node [middle] {}
      child {
        node [middle] {}
        child {
          node [middle] {}
          child { node [filled] {} }
          child { node [filled] {} }
          child { node [filled] {} }
        }
      }
    }
    child {
      node [middle] {}
      child { node [dead] {} }
      child [sibling distance = 1.5cm] {
        node [middle] {}
        child {
          node [middle] {}
          child { node [filled] {} }
          child { node [filled] {} }
        }
      }
    }
  };
\end{tikzpicture}
\caption{Schematic illustrating the general branching process of future choices. Each choice leads to an unknown number of future options to choose between, creating an expanding tree of possible future paths. The number of new options is generated from a stationary probability distribution, such that each new branch forms an independent and identically distributed tree, and the total number of options at the top of the tree is distributed according to a Galton--Watson (GW) process.  If the probability of generating no new choices is greater than zero, dead-ends can form (black circle) and there is a probability $\alpha$ that the tree will become extinct. The causal entropic collective model assumes that agents will be uniformly distributed on the final options (red circles), weighted by the probability of the tree being generated by a GW process.}
\label{fig:GW}
\end{figure}
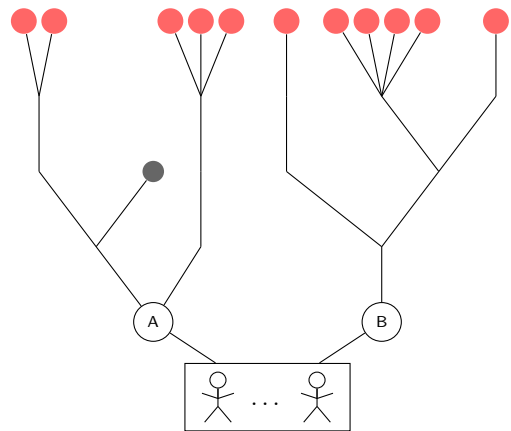
We are interested in the number of nodes on this branching tree at after some time window $h$ -- the height of the tree. The Kesten--Stigum theorem \cite{kesten1966} states that for any GW process, the distribution on the number of nodes converges to an exponential distribution, conditional on the tree not becoming extinct:
Thus, including the possibility of extinction, for large $h$, the number of options is distributed as
\begin{equation}
P(n) \simeq  \alpha \delta(0) + (1-\alpha) \frac{1}{\zeta^h} \exp \left ( -\frac{n}{\zeta^h} \right ),
\end{equation}
where $\zeta$ is the mean number of descendants of each node in each generation, $\alpha$ is the extinction probability and $\delta$ is the Dirac delta function. The extinction probability is determined by the fixed point of the generating function for the number of new choices generated on each branching. We will treat $\alpha$ as an adjustable parameter of our model. The behaviour of agents on this tree is determined, via Equation \ref{eqn:general}, by the ratio distribution $P(R) = P(\frac{n_A}{n_A + n_B})$. Since our assumption is that each new branch of the tree forms a independent GW process, this takes a simple form:
\begin{align}
P(R) &= P\bigl(n_A / (n_A + n_B)\bigr) \nonumber \\
&= \frac{1}{1+\alpha} \Bigl( (1-\alpha) + \alpha \bigl(\delta(0) + \delta(1)\bigr) \Bigr) \label{eqn:prob_r}.
\end{align}
This follows from noting that the ratio $X / (X+Y)$ of two identically distributed exponential random variables $X$ and $Y$ is a uniformly distributed random variable on (0, 1), and considering the special cases where either $n_A$ or $n_B$ is zero. Instances where both $n_A$ and $n_B$ are zero are undefined and do not contribute to the calculation. The Dirac delta functions at zero and one are the result of the possible extinction of one branch or the other. The final distribution over the choices of N agents can be obtained via Equation \ref{eqn:general}, and mirrors the distribution of $R$, with an equal probability of 1 to $N-1$ agents choosing door A, and higher probabilities for either 0 or $N$ agents to do so if $\alpha > 0$. An equivalent model exists for a tree embedded in continuous time: the Yule process \cite{yule1925amt, simon1955oac}. Thus the distribution derived for $P(R)$ does not depend on whether the branching process for possible future trees is discrete or continuous in time.


\subsection{More than two choices}
The same principles used to derive the distribution of the agents over two choices can be applied to an arbitrarily higher number of options. To do so, we need the following fact: the proportional ratios of i.i.d.\ exponential random variables ${X_1, X_2, \ldots, X_K}$ are beta distributed:
\begin{equation}
\frac{X_i}{\sum_{j=1}^K X_j} \sim \beta(1, K-1).
\end{equation}
Using this fact, along with the special cases where one or more of the trees behind each option goes extinct, we may generalise Equation \ref{eqn:prob_r}. We have the following probability distribution for $R$, the proportion of future paths behind one choice, in the case where there are three options:
\begin{align}
P(R) &=\frac{1}{1-\alpha^3} \Bigl[ \alpha(1-\alpha^2)\delta(0) + \alpha^2(1-\alpha)\delta(1) \nonumber \\
 &+{} 2\alpha(1-\alpha)^2\beta(R, 1,1) + (1-\alpha)^3\beta(R, 1,2)\Bigr], \label{eqn:prob_r_mulitple}
\end{align}
where $\beta(R; a, b)$ represents the beta probability distribution on $R$ with parameters $a$ and $b$. As $\alpha$ becomes large, the factors multiplying the beta probability distributions tend to zero faster than the delta function terms, and consensus is still enforced. Each door shares an equal probability of being the consensus choice, reflected in the greater chance that $R=0$ than $R=1$. This result mirrors the experimentally observed tendency of, for example, fish to remain as a group when presented with three options \cite{miller2013bia}, though it should be noted that this framework does not provide a clear way to model groups with conflicting preferences -- a limitation addressed in the discussion. For all but the highest values of $\alpha$, the probability of a consensus decreases with the number of options $M$, implying the common-sense notion that the probability of all agents choosing the same option is reduced as the number of equivalent choices becomes very high.

\section{Consequences}
\subsection{Consensus decision making}
The causal entropic model predicts a tendency for groups of agents to reach a consensus. In the case where the extinction probability is greater than zero, there is a strong entropic `bonus' for agents to remain as a single group, specified by the Dirac delta functions in Equations \ref{eqn:prob_r} and \ref{eqn:prob_r_mulitple}. However, even in the case where the probability of all but one future tree becoming extinct is effectively zero, such as when $\alpha$ is zero or very small, or when the number of choices is very high, there is still a strong bias towards consensus decisions. For example, in the case that $\alpha = 0$, in Figure 4 we plot the probability that a group of agents of size 2, 3 or 4 will choose the same option from $K$ independent choices, compared to the probability of this occurring if each agent makes a choice uniformly at random.

\begin{figure}
\centering
\psff{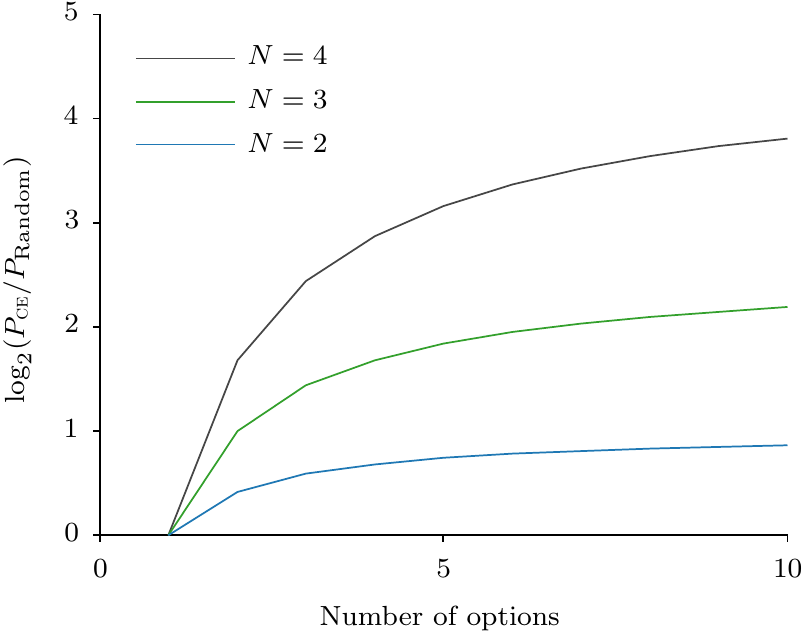}
\caption{The log ratio of the probability that a group of $N$ agents make the same choice from $K$ independent options within the causal entropic (CE) model, relative to random chance. The ratio is always above one for $K>1$ and increases with both $K$ and $N$, indicating the causal entropic model's bias towards consensus collective decision-making.}
\label{fig:consensus}
\end{figure}

\subsection{Social interaction rules}
The entropic prediction of collective consensus is fundamentally a group-level analysis. Most studies in collective decision making have started from a model of how individuals react to the decisions of others. What individual interaction rules would be necessary to produce the group-level behaviour that our analysis predicts? We can answer this question by considering a single individual choosing from two options when the other members of the group have already decided. We simply reduce the number of degrees of freedom in Equation \ref{eqn:general} to this single individual, assuming that $A$ and $B$ agents have already assigned themselves to options A and B:
\begin{equation}
P(\textrm {A} \mid A, B) = \int_{{R}=0}^1 P(R) R^{A+1} (1-R)^{B} \, \mathrm{d}R.
\end{equation}
The result of this calculation for an example case where $\alpha = 0$ is shown in Figure 5 (a). By inspection, the probability of choosing A dictated by this rule looks like a Weber's law \cite{weber1834dpr} -- choosing A in proportion to $(A + \varepsilon)/(A+B + 2\varepsilon)$. To test the similarity of our result to this rule, we plot the probability of choosing A against $(A +1)/(A+B+2)$ in Figure 5 (b) and find a perfect match. As $\alpha$ increases the value of $\varepsilon$ decreases, so for $\alpha \simeq 1$, $\varepsilon \simeq 0$. A brief consideration of this rule in the case of high $\alpha$ shows that it enforces the same consensus as derived at the group level, since the first agent to commit to option A or B makes the probability of that option for subsequent agents approximately equal to one, thus causing an irreversible information cascade.

\begin{figure}
\centering
\psff{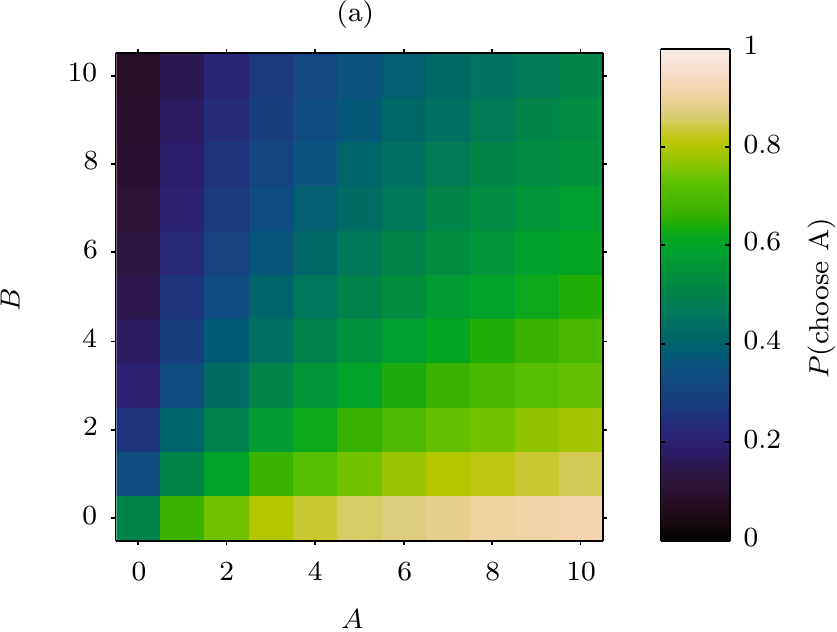}
\psff{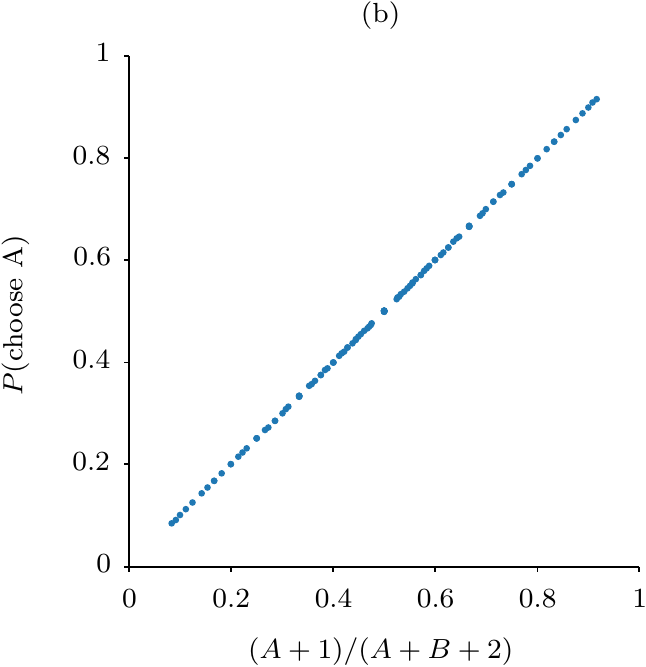}
\caption{The social interaction rule implied by the causal entropic model. Panel A shows the probability one individual will choose option A rather than B, conditioned on $A$ and $B$ agents already committing to each option. The resulting probability is a match to Weber's law with one `pseudo-observation', $P(\textrm{A} \mid A, B) = (A+1)/(A+B+2)$, as shown in panel B.}
\label{fig:social_decision_rule}
\end{figure}

\subsection{Collective intelligence}
The entropic enforcement of consensus decisions implies some degree of collective intelligence. To see this, consider the model used by Ward \emph{et al.}\ \cite{ward2011faa} to explain the collective decisions of groups of varying size. In this 'Many-Eyes' model, if any one agent in a group spots a threat, all agents will avoid it. This implies that the proportion of agents avoiding a threat should grow in proportion to the probability that at least one will spot the threat, i.e., $1-0.5(1-d)^N$, where $d$ is the detection probability.

Our model implies a similar result. Since the agents themselves are not {\it actively} trying to maximise entropy (instead, social decision rules have evolved that tend to maximise entropy in general), any agent seeing a threat should avoid it. However, once this occurs, the general tendency of the other agents to maintain a consensus means that the group will generally stick together, with a probability determined by the extinction probability of the branching process, closely mimicking the many-eyes model. We can calculate the expected number of agents avoiding a threat as a function of the extinction probability, by conditioning Equation \ref{eqn:general} on a given number, $N_d$, detecting the threat and avoiding it, and weighting by the probability of that number of detections, given a detection probability $d$.
\begin{multline}
P(A \, \textrm{avoid threat}) = \sum_{i=0}^{N}  \Biggl[ \binom{N}{i} d^i (1-d)^{N-i} \\
 \times \binom{N}{A-i} \int_{R=0}^1 P(R) R^{A-i} (1-R)^{N-A-i} \, \mathrm{d}R \Biggr].
\end{multline}
In Figure 6 we plot the implied collective intelligence for different values of the extinction probability, in the case where any given agent has a $d=0.1$ chance of detecting a hidden threat, as in the example of Ward \emph{et al.} \cite{ward2011faa}. The prediction for high values of $\alpha$ is essentially identical to the prediction of \cite{ward2011faa}, in that consensus is entirely enforced.
\begin{figure}
\centering
\psff{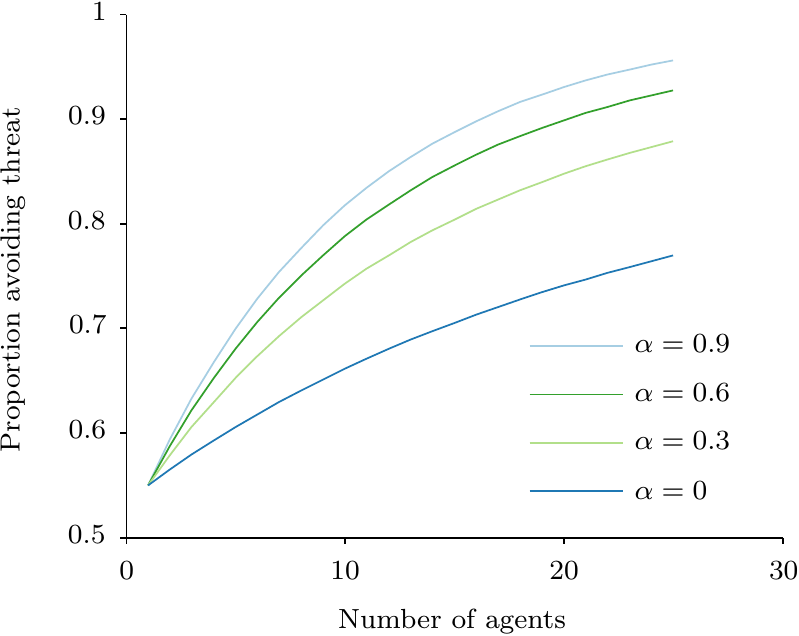}
\caption{The increasing proportion of agents avoiding a threat with a low detection probability ($d=0.1$) as a function of group size, for different values of the extinction probability on the future paths tree. The greater the possibility of one or other of the future path trees becoming extinct, the greater the cohesive force between the agents, and thus the stronger the information transfer between the detecting agents and the others, resulting in improved collective intelligence.}
\label{fig:swarm_intelligence}
\end{figure}

\subsection{Collective motion: derivation of a social force}
Set in a discrete space, the model we describe here does not give immediate quantitative predictions about the types of collective motion \cite{vicsek2012cm} we would expect in a continuous space. However, we can sketch what such a generalised model would look like. The choice of which direction to move in is a decision like any other, but with many possible options. According to the arguments above, agents in approximately the same spatial location -- those who will experience the same branching process of future options based on their choice -- should tend to move in the same direction. 

We can imagine that for a given type of agent there is a typical spatial range over which future trees are strongly correlated. We can associate this with the zones of interaction found in many models of collective motion, such as the classic Vicsek model \cite{vicsek1995nto} and Couzin zonal model \cite{couzin2002cma}. In the case of a relatively confined environment, individuals outside of the immediate perceptual range may still experience the same future trees, and this can be expressed in individual interaction terms via a memory of encounters \cite{mann2013msi}, leading to something akin to a mean-field model. As for the first question, what interaction rule individuals use to align their movements, this can be answered at least in general terms from the discussion of social decision rules above. The interaction must be some form of Weber's law, with the probability to adopt a given direction varying with the proportion of individuals within the interaction zone who have that direction. This is very similar to the rules that are actually employed in such individual-level models. For example, in the Couzin model, the alignment force on an agent along any axis is proportional to the sum of each other individual's motion along that axis, divided by the total number of individuals.

We provide a quantitative special case. Assume that in a continuous space there are $N+1$ agents, of which $N$ are already committed to a particular position. Where should the $(N+1)$th agent position itself? At each point $x_i$ in the space-direction continuum occupied by an agent in $\{1, \ldots, N \}$, there is a probability distribution over the ratio $P(\frac{n_i}{\sum_j n_j} \mid x_i)$, which we will assume is equal for all points $x_i$ (we will be ignoring second-order effects where these $N$ agents are influencing each other, and focus only on the $(N+1)$th agent). Let us further assume a particular form for the distribution $P(\frac{n_{N+1}}{\sum_j n_j} \mid x_{N+1})$: with probability $\beta_k$, this position shares a future tree with position $x_k$. We take this probability to be defined by a squared exponential decay function:
\begin{equation}
\beta_k = \exp \left ( \frac{-(x_{N+1}-x_k)^2}{L} \right ).
\end{equation}
The probability distribution of possible position choices $x_{N+1}$ is determined by a mixture of the possibility that $P(n_{N+1} / \sum_j n_j)$ is independent of all other points, and each of the possibilities that $x_{N+1}$ shares a future tree with the position of another agent. Again ignoring second-order effects, we have
\begin{align}
P(x_{N+1}) &\simeq (1-\sum_k \beta_k) \left\langle\frac{n_{N+1}}{\sum_j n_j}\right\rangle \prod_i \left\langle\frac{n_i}{\sum_j n_j}\right\rangle \nonumber \\
&\mspace{50mu} {} +  \sum_{k=1}^N\beta_k \left\langle \biggl( \frac{n_k} {\sum_j n_j} \biggr)^2 \right\rangle \prod_{i \neq k} \left\langle\frac{n_i}{\sum_j n_j}\right\rangle \nonumber \\
&= \mu^{N+1} +  \sum_{k=1}^N \beta_k \mu^{N-1} \sigma^2,
\end{align}
where $\mu =  \left\langle \frac{n_i}{\sum_j n_j} \right\rangle$ and $\sigma^2 = \left\langle \left(\frac{n_i}{\sum_j n_j}\right)^2 \right\rangle - \mu^2$. We find the optimal position by maximising $P(x_{N+1})$, obtaining:
\begin{equation}
\frac{d}{dx_{N+1}}  \sum_{k=1}^N \exp \left ( \frac{-(x_{N+1}-x_k)^2}{L} \right ) = 0,
\end{equation}
which we identify as the least-squares solution: $x_{N+1} = \frac{1}{N} \sum_{k=1}^N x_k$. Therefore the unique optimal position for agent $N+1$ is the mean position of all the other agents, implying a social `force' towards this point proportional to Equation \ref{eqn:social_force}. It should be clear that the same argument would apply in relation to the direction choices of other individuals as well, creating an equivalent `social force' to rotate the agent's direction towards the average of the group. If the probability of sharing future trees were correlated between space and direction, then a distance-dependent alignment force would emerge.

%

\section{Discussion}
We have demonstrated that the causal entropic principle gives a purely statistical prediction for many of the emergent properties of collective behaviour, without any detailed understanding on the mechanisms of interactions between individuals. Adopting the taxonomy of modelling approaches described by Sumpter \emph{et al.}\ \cite{sumpter2012tmc}, this is a purely global approach to modelling groups and is complementary to a detailed understanding of individual behaviour, rather than a replacement. Our model takes an entirely novel approach to understanding the origins of collective behaviour, and makes testable predictions about the fundamental form of social interactions

Our model predicts that interactions between individuals take the form of a Weber's law. This social decision rule has empirical support in the response to various stimuli of several species, e.g., insects \cite{smith1994nop, perna2012irf}, fish \cite{arganda2013acr}, and humans \cite{deco2007wli, mann2013tdo, spaiserschools}, as well as a solid grounding in the psychophysics of estimating differences \cite{stevens1957otp, johnson2002brn, dehaene2003tnb}. However, there are other studies that find that individual decisions are better described by more non-linear interactions e.g.\ \cite{sumpter2008cdm, Ward2008ly}. Perna \emph{et al.}\ \cite{perna2012irf} have already shown that an accumulation of Weber's law interactions, combined with some degree of noise or inaccuracy (which we would expect in any real system) can lead to apparently non-linear interactions. We therefore suggest that where non-linear interactions are observed, these may be the result of an accumulation of smaller-scale Weber interactions. In many experimental setups involving animal groups, the choices ultimately made by the individuals are not single events, but the final result of a period of motion where many smaller choices are made, which supports the idea that the final choice can be seen as an accumulation of smaller interactions. Simulations show that the type of linear, Weber-type interactions used in self-propelled particle models can lead to strongly non-linear consensus decision making in moving groups \cite{couzin2005ela, Leonard2012dvc}.

A limitation of our model is the lack of a description of groups with conflicting information or preferences. Variation in information or personality in groups has been shown to be a potentially important driver of collective outcomes \cite{couzin2005ela, couzin2011uip, aplin2014ilp}. This could potentially be addressed by assigning different beliefs to each agent about the probability distribution on future trees. However, we have deliberately framed our model in terms of a consistent rule that produces a maximum entropy result over all possible futures, rather than assigning entropy-maximising agency to the individuals themselves. There is no clear reason for individual agents, animal or human, to \emph{desire} greater entropy over future paths; rather, we consider it as a result of uncertainty in which futures may be possible, and which decisions the agents will take. Nonetheless, the viewpoint could be relaxed to allow the emergence of a more sophisticated model including conflicting groups in the future. The entropic consequences of conflict are therefore an area of importance for future research in this area.

The model described in this paper gives a simple caricature of the types of decisions that face groups of intelligent agents. This abstraction is useful for understanding the logic of how causal entropic maximisation implies group behaviours, social interactions, and collective intelligence. We have shown how the model might be generalised to a continuous space in consideration of collective motion. Such an expansion of the model could potentially describe the structure of moving animal groups \cite{ballerini2008ira, ballerini2008eio} and patterns of group-level direction changes \cite{Buhl:2006ys}. More widely, the causal entropic principle may provide a general framework for understanding the dynamics of complex intelligent systems, extending from animal groups, through organisations such as corporations and governments, to global human social systems built on the enormous connectivity of the internet. We cannot be sure what series of choices every animal, pedestrian, car driver, bureaucrat, or social-network user will make over an extended period of time. But precisely this ignorance can help us to predict what they will do next.

\begin{acknowledgments}
RPM was supported by ERC Starting Grant `IDCAB' reference number 220/104702003 to David JT Sumpter and ERC Advanced Investigator Grant `Momentum' reference number 324247 to Dirk Helbing. RG was supported by the German Science Foundation (DFG)
under the reference number GA 1615/1-1.
\end{acknowledgments}

\bibliographystyle{ieeetr}
\bibliography{pigeon}

\end{article}

\end{document}